\documentclass[runningheads]{llncs}
\usepackage[T1]{fontenc}
\usepackage{graphicx}
\usepackage[table,xcdraw]{xcolor}
\usepackage{amsfonts} 
\usepackage{float}
\usepackage{color,soul}
\usepackage[sorting=none]{biblatex}
\bibliography{mybibliography}
\begin{document}
\title{Morphology-preserving Autoregressive 3D Generative Modelling of the Brain}
\titlerunning{Morphology-preserving 3D Generative Modelling of the Brain}
%
\author{
    Petru-Daniel Tudosiu$^{@*}$\inst{1} \orcidID{0000-0001-6435-5079} \and
    Walter Hugo Lopez Pinaya \inst{1} \orcidID{0000-0002-2795-9209} \and
    Mark S. Graham \inst{1} \orcidID{0000-0002-4170-1095} \and
    Pedro Borges \inst{1} \orcidID{0000-0001-5357-1673} \and
    Virginia Fernandez \inst{1} \orcidID{0000-0001-5984-197X} \and
    Dai Yang \inst{2} \and
    Jeremy Appleyard \inst{2} \and
    Guido Novati$^{\#}$ \inst{3} \and
    Disha Mehra \inst{2} \and
    Mike Vella$^{\#}$ \inst{4} \and
    Parashkev Nachev \inst{5} \orcidID{0000-0002-2718-4423} \and
    Sebastien Ourselin \inst{1} \orcidID{0000-0002-5694-5340}\and
    Jorge Cardoso \inst{1} \orcidID{0000-0003-1284-2558}
}
\authorrunning{P.-D. Tudosiu, W.H.L. Pinaya et al.}

\institute{
Department of Biomedical Engineering, School of Biomedical Engineering $\&$ Imaging Sciences, King’s College London, London, UK \and
NVIDIA \and
DeepMind \and
Oxford Nanopore Technologies, Gosling Building, Oxford Science Park, Edmund Halley Rd, Littlemore, Oxford OX4 4DQ, UK \and
Queen Square Institute of Neurology, University College London, London, UK \\
$^{@}$ Corresponding Author \\
$^{\#}$ Work done while at NVIDIA
}

\maketitle              
\begin{abstract}
Human anatomy, morphology, and associated diseases can be studied using medical imaging data. However, access to medical imaging data is restricted by governance and privacy concerns, data ownership, and the cost of acquisition, thus limiting our ability to understand the human body. A possible solution to this issue is the creation of a model able to learn and then generate synthetic images of the human body conditioned on specific characteristics of relevance (e.g., age, sex, and disease status). Deep generative models, in the form of neural networks, have been recently used to create synthetic 2D images of natural scenes. Still, the ability to produce high-resolution 3D volumetric imaging data with correct anatomical morphology has been hampered by data scarcity and algorithmic and computational limitations. This work proposes a generative model that can be scaled to produce anatomically correct, high-resolution, and realistic images of the human brain, with the necessary quality to allow further downstream analyses. The ability to generate a potentially unlimited amount of data not only enables large-scale studies of human anatomy and pathology without jeopardizing patient privacy, but also significantly advances research in the field of anomaly detection, modality synthesis, learning under limited data, and fair and ethical AI. Code and trained models are available at: https://github.com/AmigoLab/SynthAnatomy.
\end{abstract}
\keywords{Transformers \and VQ-VAE \and Generative Modelling \and Neuroimaging \and Neuromorphology.}

\section{Introduction}
Current advances in the application of deep learning (DL) in medical imaging were driven by substantial initiatives and challenges such as UK Biobank (UKB)\cite{ukb}, Alzheimer’s Disease Neuroimaging Initiative (ADNI)\cite{adni}, and the Medical Segmentation Decathlon\cite{decathlon}. However, these are relatively small compared to computer vision datasets. Owing to the lack of access to sufficient data due to privacy concerns, medical imaging data is not fully leveraging DL's full potential and this hinders its translation from research to the clinical environment. State-of-the-art (SOTA) algorithms rely on a handful of highly curated datasets which could lead to biases due to imbalanced demographics or acquisition parameters, that may negatively affect their performance for certain populations. A solution to this problem could come from the generative modelling of the underlying available data to balance the prevalence of confounding variables in the training dataset.

While semi-supervised 3D generative modelling of the brain has been steadily explored and improved \cite{supervised_synthesis_1,supervised_synthesis_2,supervised_synthesis_3}, progress in unsupervised generative modelling has been more limited. Generative Adversarial Network (GAN) based approaches, which suffer from memory constraints and stability issues, have mostly been trained on low-resolution 3D images \cite{unsupervised_synthesis_1,kwon_brain,xing_brain}, having only recently been able to synthesise full resolution images via learning partial sub-volumes \cite{sun_brain}. Whereas previous methods quantify sample diversity using classic metrics such as Multi-Scale Structural Similarity Index (MS-SSIM) \cite{msssim}, distribution alignment via Fr\'echet Inception Distance (FID) \cite{fid} and Maximum Mean Discrepancy (MMD) \cite{mmd}, none have quantified if the generated data preserves the morphological characteristics of the data -- crucial if we are to use such methods. 

Recently, autoregressive models have achieved SOTA results synthesising high resolution natural images \cite{vqvae2,taming_transformer,vqvae_transformer_1}. This was accomplished by employing a compression model, namely a Vector Quantised-Variational Autoencoder (VQ-VAE) \cite{vqvae1, vqvae2}, to project the images into a discrete latent representation where the images' likelihood becomes tractable. An attention-based Transformer network \cite{transformer, performer} is then used to model the product of conditional distributions by maximising the expected log-likelihood of the training data.

Following \cite{sdd} as part of the Synthetic Data Desiderata, a good synthetic dataset should share many if not all statistical properties of the real dataset. One such property, if not the most important, of synthetic structural medical images is their morphological correctness. Covariates of interest such as demographic and pathological ones determine the phenotype of each subject which in turn contributes to the population-level morphological statistics. Without it, any development done on the synthetic data as part of the Train on Synthetic, Test on Real \cite{tstr} paradigm could suffer from higher domain distribution shifts slowing down the development. Furthermore, without morphological assessment, any hypothesis tested on the synthetic data would be rendered highly uncertain.

In this study, we scale and optimise VQ-VAE and Transformer models for high-resolution volumetric data, aiming to learn the data distribution of both radiologically healthy and pathological brains. A thorough morphological evaluation is employed by using Voxel-Based Morphometry (VBM) \cite{vbm} and volumetric analysis using Geodesic Information Flows (GIF) \cite{gif}, demonstrating that synthetic data generated by the proposed model preserves the morphological characteristics and phenotype of the data.

\section{Background}

Our model is based on the two-stage architecture introduced by \cite{vqvae1,vqvae2} and extended by \cite{taming_transformer}, where a VQ-VAE model is used to project a high-resolution image into a compressed latent representation and a transformer is trained to maximize the likelihood of the flattened representations. 

\subsection{VQ-VAE}

The VQ-VAE \cite{vqvae1, vqvae2} is comprised of an encoder $E$ that projects the input image $\mathbf{x}\in{\mathbb{R}}^{H \times W \times D}$ to a latent representation space $\widehat{\mathbf{z}} \in \mathbb{R}^{h \times w \times d \times {n}_{z}}$ where ${n}_{z}$ is the latent embedding vector's dimensionality. Afterwards, an element-wise quantization is done for each spatial code $\widehat{\mathbf{z}}_{ijk}\in\mathbb{R}^{n_z}$ onto its nearest vector $e_k\in\mathbb{R}^{n_z}, k \in 1,..., K$ from a codebook, where $K$ denotes the vocabulary size of the codebook, obtaining $\widehat{\mathbf{z}}_{q}$. The codebook's elements are learned in an online manner, together with the other model's parameters. Based on the quantized latent space, a decoder $G$ tries to reconstruct the observations $\widehat{\mathbf{x}}\in\mathbb{R}^{H \times W \times D}$. By replacing each of the codebook elements vector $\widehat{\mathbf{z}}_{q} \in \mathbb{R}^{h \times w \times d \times n_z}$ with their associated index $k$, the latent discrete representation is obtained.

\subsection{Transformer}

Transformers models and their associated self-attention mechanisms can capture the interactions between inputs regardless of their relative positioning. Due to this, the attention mechanism scales quadratically with the size of the input sequence. Since the VQ-VAE's latent discrete representation when applied to volumetric medical data is 3D and thus large in scale, standard transformers do not scale to the necessary sequence length. Recently, multiple advances have made Transformers more efficient \cite{linear_transformers}; models such as the Performer, with its FAVOR+ linear scaling attention approximation \cite{performer} offers a good compromise between accurately modelling long-sequences while preserving a reasonable computational complexity \cite{linear_transformers}. Thus the Performer is used to model the latent sequences; by minimizing the conditional distribution of codebook indices $p(s_i)=p(s_i|s_{<i})$ on the flattened 1D sequences of the 3D latent discrete representations, the data log-likelihood is maximized in an autoregressive fashion.

\section{Methods}
\subsection{Descriptive Quantization for Transformer Usage}

To create a Transformer-based generative model of the brain, the image volume needs to be transformed into a 1D sequence of tokens. To achieve this, a VQ-VAE model that reduces the overall spatial size by a factor of 4096, allowing an input image of size X to be represented by a sequence of 1400 tokens. This 1400-long token sequence is learnt in an online fashion together with the VQ-VAE model by using the Exponential Moving Average (EMA) algorithm \cite{vqvae1,vqvae2} as per Eq. \ref{eq:non_ema_vq}.
\begin{equation}
    \resizebox{0.9\columnwidth}{!}{$%
        {\mathcal{L}}_{VQ-VAE}(\mathbf{x}, G(\mathbf{\widehat{\mathbf{z}}_{q}}))={\mathcal{L}}_{Rec} + {\mathcal{L}}_{Adv} +\|s g[E(\mathbf{x})]-\mathbf{\widehat{\mathbf{z}}_{q}}\|_{2}^{2}+\beta\|s g[\mathbf{\widehat{\mathbf{z}}_{q}}]-E(\mathbf{x})\|_{2}^{2}$%
    }%
\label{eq:non_ema_vq}
\end{equation}
\begin{equation}
    \resizebox{0.9\columnwidth}{!}{$%
        N_{i}^{(t)}:=N_{i}^{(t-1)} * \gamma+n_{i}^{(t)}(1-\gamma), \quad m_{i}^{(t)}:=m_{i}^{(t-1)} * \gamma+\sum_{j}^{n_{i}^{(t)}} E(x)_{i, j}^{(t)}(1-\gamma), \quad \widehat{\mathbf{z}}_{qi}^{(t)}:=\frac{m_{i}^{(t)}}{N_{i}^{(t)}}$%
    }%
\label{eq:ema_vq}
\end{equation}

\noindent where $sg$ stands for the stop-gradient operation. As per \cite{vqvae1,vqvae2}, the third loss component in Eq. \ref{eq:non_ema_vq} is replaced by Eq. \ref{eq:ema_vq}, where $n_{i}^{(t)}$ stands for the number of vectors in $E(\mathbf{x})$ that will be quantized to codebook element $\widehat{\mathbf{z}}_{qi}$. The hyper-parameters $\gamma$ and $\beta$ control the decay of the EMA and the commitment of the encoder output to a certain quantized element respectively. 

For the codebook to be perceptually rich, a loss similar to \cite{taming_transformer, vqgan_loss} which is formed by the first and second elements of Eq. \ref{eq:non_ema_vq} as defined bellow:
\begin{equation}
    \resizebox{0.9\columnwidth}{!}{$%
    {\mathcal{L}}_{Rec} = \|\mathbf{x}-\widehat{\mathbf{x}}\|_{1} + \||FFT(\mathbf{x})|-|FFT(\widehat{\mathbf{x}})|\|_{2} + {\mathcal{LPIPS}}_{0.5}(\mathbf{x},\widehat{\mathbf{x}})$%
    }%
    \label{eq:vqvae_reconstruction}
\end{equation}

Where the first term is a pixel-space L1 norm, the second term is the L2 norm of the image's Fourier representations based on \cite{jukebox} which aims at facilitating high-frequency feature preservation, the third term is the LPIPS \cite{lipis} loss using AlexNet applied on 50\% of slices on each axis. Lastly, the ${\mathcal{L}}_{Adv}$ is based on a Patch-GAN discriminator-based adversarial loss \cite{patchgan,taming_transformer}, replacing the original loss by the LS-GAN \cite{lsgan} one (see Eq \ref{eq:ls_gan}):
\begin{equation}
    \resizebox{0.9\columnwidth}{!}{$%
    \begin{array}{l}
        \min _{D} {\mathcal{L}}_{LSGAN}(D)=\frac{1}{2} \mathbb{E}_{\mathbf{x} \sim p_{\mathrm{data}}(\mathbf{x})}\left[(D(\mathbf{x})-1)^{2}\right]+\frac{1}{2} \mathbb{E}_{\mathbf{x} \sim p_{\mathbf{data}}(\mathbf{x})}\left[(D(G(\widehat{\mathbf{x}})))^{2}\right] \\
        \\
        \min _{G} {\mathcal{L}}_{LSGAN}(G)=\frac{1}{2} \mathbb{E}_{\mathbf{x} \sim p_{\mathbf{x}}(\mathbf{z})}\left[(D(G(\widehat{\mathbf{x}}))-1)^{2}\right]
    \end{array}$%
    }%
    \label{eq:ls_gan}
\end{equation}

Each of these losses independently contributes to model training stability and reconstruction quality.

\subsection{Autoregressive Modelling of the Brain}

The VQ-VAE model was first trained on T1w MRI images of neurologically healthy subjects from UKB \cite{ukb} until convergence, and then their $z_q$ representations were extracted. Afterwards, further fine-tuning on the pathological dataset formed from the baseline T1w MRI scans of ADNI \cite{adni} subjects was done until over-fitting was noticed, at which point the ADNI subjects' $z_q$ representation was also extracted. This paradigm was chosen since in \cite{ukbadni} it was shown to either be on par or better compared to training a VQ-VAE model only on the pathological dataset. Furthermore, we aim also to highlight that the pre-trained model can be fine-tuned and learn new morphology, in this case, a pathological one, thus increasing the usefulness of the UKB trained VQ-VAE as a pretrained model for the community.

In order to ensure a higher quality of the Transformer's samples, the top 1\% generated samples, based on the score obtained by averaging the Patch-GAN discriminator output, were used in this work.

As the VQ-VAE representations cover all phenotypes, a separate Transformer model has been trained on the latent representations of different sub-populations to model individual morphological subgroups. More specifically, to demonstrate the morphological phenotype preservation, the UKB \cite{ukb} dataset was partitioned into young vs. old sub-populations, and small vs. big ventricles sub-populations. We defined all of these groups based on the first and last of five quantiles based on "age when attended assessment centre" (21003-2.0) and "volume of the ventricular cerebrospinal fluid" (25004-2.0) UKB variables, respectively. To test the preservation of disease morphology, we split the ADNI dataset into cognitively normal (CN) and Alzheimer's disease (AD) subgroups based on the "diagnosis/scan category assignment field".

\section{Experiments and Results}
The performance of the proposed model is assessed in two ways: first, the quality of generated samples is measured according to image fidelity metrics commonly used in generative models; second, we verify if the morphological characteristics of a population and the differences between sub-populations are preserved when comparing real and synthetic data. We compared our model to a baseline volumetric VAE model. The models by \cite{kwon_brain,xing_brain, wgan, lsgan} underwent extensive hyper-parameter exploration at the original resolutions but failed to converge on our data. Only the VAE results are thus presented as a baseline.   

\subsection{Quantitative Image Fidelity Evaluation}

\begin{figure}[b!]
\centering
\includegraphics[width=1\columnwidth]{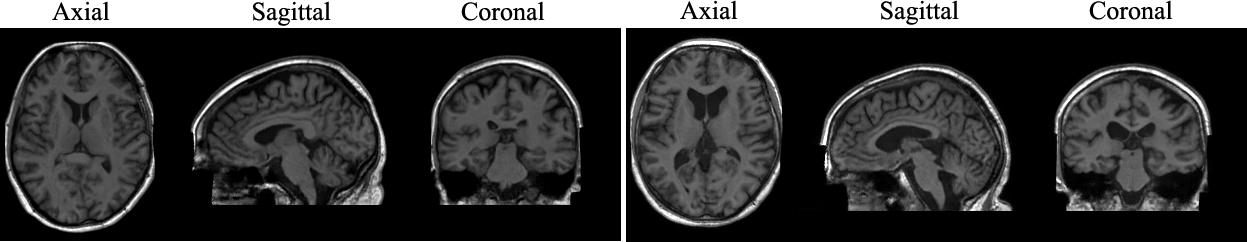}
\caption{Synthetic samples. On the left UKB small ventricles and on the right UKB big ventricles.} 
\label{fig:syn}
\end{figure}

Similarly to \cite{kwon_brain,xing_brain}, we use the FID \cite{fid} to assess the visual quality of the generated images. Since originally the metric is based on a pretrained Inception V3 network on 2D natural images, it cannot be applied on 3D volumes directly, so here it is applied on the middle slice of each axis and reported individually. To measure the quality of the 3D samples, batch-wise MMD with a dot product as the kernel is being used as suggested in \cite{kwon_brain,xing_brain}. Briefly, MMD quantifies the distance between the distributions with finite sample estimates in kernel functions in the reproducing kernel Hilbert space \cite{mmd}. Lastly, to estimate the diversity of the generated images, MS-SSIM is being used in a pair-wise fashion between the generated synthetic samples as in \cite{kwon_brain,xing_brain}. For easy comparison, all metrics have also been calculated between each sub-population's real images such that a ground truth baseline is also offered.

\begin{table}[b!]
\centering
\resizebox{1\columnwidth}{!}{%
\begin{tabular}{cccccc}
Model         & Dataset & Population           & FID (Ax|Cor|Sag)         & ${bMMD}^{2}$           & MS-SSIM                   \\ \hline
Real          & UKB     & Young                & 0.35 | 0.85 | 0.42       & $0.00208_{\pm0.00026}$ & $0.65_{\pm0.08}$          \\
VQ-VAE (ours) & UKB     & Young                & 31.04 | 57.19 | 57.19    & $0.00903_{\pm0.00090}$ & $0.68_{\pm0.03}$          \\
VAE           & UKB     & Young                & 193.56 | 302.74 | 251.34 & $0.02757_{\pm0.00091}$ & $0.15_{\pm0.001}$         \\ \hline
Real          & UKB     & Old                  & 1.16 | 1.42 | 0.37       & $0.00217_{\pm0.00045}$ & $0.65_{\pm0.07}$          \\
VQ-VAE (ours) & UKB     & Old                  & 33.68 | 60.60 | 78.82    & $0.00887_{\pm0.00104}$ & $\mathbf{0.67_{\pm0.03}}$ \\
VAE           & UKB     & Old                  & 234.86 | 289.21 | 242.18 & $0.02622_{\pm0.00044}$ & $0.15_{\pm0.001}$         \\ \hline
Real          & UKB     & Small Ventricles     & 1.74 | 1.99 | 0.87       & $0.00220_{\pm0.00044}$ & $0.67_{\pm0.07}$          \\
VQ-VAE (ours) & UKB     & Small Ventricles     & 28.33 | 58.23 | 76.68    & $0.00892_{\pm0.00106}$ & $\mathbf{0.70_{\pm0.04}}$ \\
VAE           & UKB     & Small Ventricles     & 206.92 | 318.37 | 258.17 & $0.02836_{\pm0.00078}$ & $0.14_{\pm0.001}$         \\ \hline
Real          & UKB     & Big Ventricles       & 1.15 | 1.44 | 0.53       & $0.00231_{\pm0.00041}$ & $0.64_{\pm0.05}$          \\
VQ-VAE (ours) & UKB     & Big Ventricles       & 36.02 | 57.76 | 76.51    & $0.00937_{\pm0.00069}$ & $0.68_{\pm0.04}$          \\
VAE           & UKB     & Big Ventricles       & 215.37 | 293.97 | 244.84 & $0.02738_{\pm0.00058}$ & $0.16_{\pm0.001}$         \\ \hline
Real          & ADNI    & Cognitively Normal   & 21.49 | 17.31 | 9.34     & $0.00123_{\pm0.00021}$ & $0.56_{\pm0.05}$          \\
VQ-VAE (ours) & ADNI    & Cognitively Normal   & 53.88 | 93.62 | 112.32   & $0.01558_{\pm0.00348}$ & $0.71_{\pm0.08}$          \\
VAE           & ADNI    & Cognitively Normal   & 233.59 | 397.52 | 421.04 & $0.02562_{\pm0.00119}$ & $0.14_{\pm0.05}$          \\ \hline
Real          & ADNI    & Alzheimer's Diseased & 9.08 | 16.85 | 13.49     & $0.00167_{\pm0.00034}$ & $0.55_{\pm0.13}$          \\
VQ-VAE (ours) & ADNI    & Alzheimer's Diseased & 87.75 | 51.74 | 90.95    & $0.01562_{\pm0.00304}$ & $0.61_{\pm0.11}$          \\
VAE           & ADNI    & Alzheimer's Diseased & 235.33 | 332.70 | 340.78 & $0.02804_{\pm0.00177}$ & $0.12_{\pm0.06}$          \\ \hline
\end{tabular}%
}
\caption{The ${bMMD}^2$ and MS-SSIM were calculated on 3D generated images while FID was done middle-slices-wise of generated volumes.}
\label{tab:qualitative_metrics}
\end{table}
Across the board, as is described in Table \ref{tab:qualitative_metrics}, both in regards to the sub-populations and axial, coronal and sagittal slices, the FID of the VQ-VAE model outperforms the VAE baseline by a high margin, showcasing the realistic appearance of the sampled synthetic brains as seen in Fig. \ref{fig:syn}. The same can be said for the $bMMD^2$, where the VQ-VAE is one order of magnitude smaller for UKB sub-populations and substantially better for the ADNI sub-populations. The difference in $bMMD^2$ performance between VQ-VAE's UKB and ADNI sub-populations might be because the ADNI dataset is considerably smaller than the UKB one, and to circumvent that, the VQ-VAE compression model was firstly trained on the UKB dataset and then fine-tuned on the ADNI one. Thus, the $z_q$ representation fed into the Transformer, which is the generative model per se, is not specialised for ADNI, but instead, it tries to encompass it. Finally, MS-SSIM shows that the VQ-VAE achieves a life-like high diversity of samples across all sub-populations, significantly surpassing the VAE. The peculiar case of the ADNI AD sub-population might be attributed to the same cause as the $bMMD^2$.

\subsection{Morphological Evaluation}
\begin{figure}[b!]
\centering
\includegraphics[width=0.8\columnwidth]{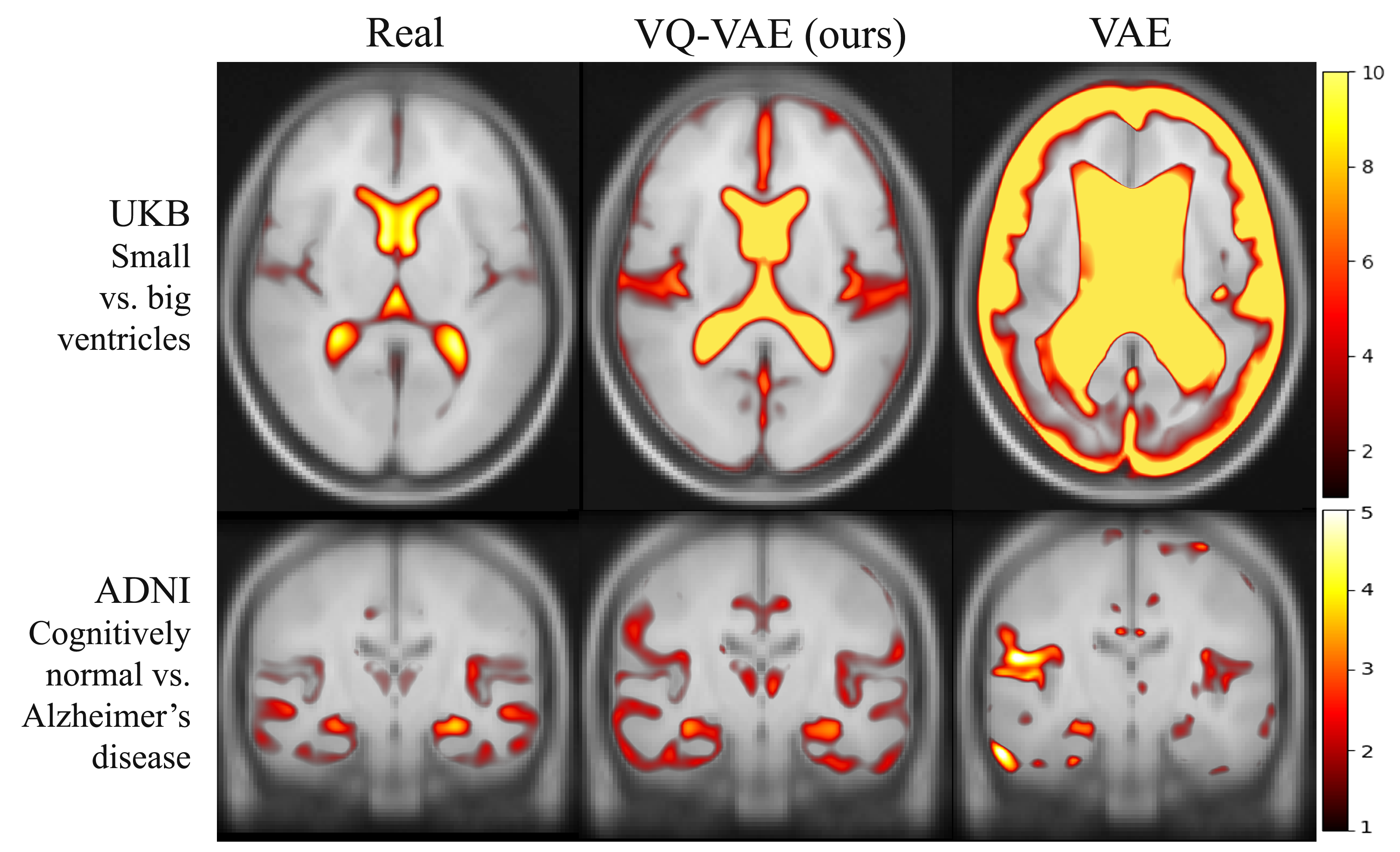}
\caption{Thresholded uncorrected VBM t-statistics maps processed as per \cite{vbm_hat} showcasing the morphological differences between two populations based on real samples, VQ-VAE synthetic samples, and VAE synthetic samples. For UKB small vs. big ventricles modulated CSF tissue segments were used, while for ADNI, cognitively normal vs. AD modulated GM tissue segments were used.} 
\label{fig:vbm}
\end{figure}

To evaluate the morphological correctness of the synthetic samples, Voxel-Based Morphometry (VBM) \cite{vbm} was used to investigate the focal differences in the brain anatomy of the sub-populations. At the core of VBM stands the application of a generalised linear model and associated statistical tests across all voxels of a group-aligned population, to identify morphological differences in modulated tissue compartment between the selected groups.

The VBM analysis did not factor out any covariates available in the real datasets since the generative process was unconditioned. All t-statistics maps have been corrected to minimise the effects of low variance areas following \cite{vbm_hat}. As shown in Fig. \ref{fig:vbm} the t-statistics maps between synthetic images generated by the VQ-VAE strongly agree with the VBM maps of real data, primarily when compared with the VAE baseline. In the UKB small ventricles vs. big ventricles experiment, VQ-VAE again successfully models the ventricular differences correctly compared to the real-data VBM maps, while the VAE model strongly emphasizes them or exacerbates the subarachnoid CSF. Lastly, the morphological differences between cognitively normal vs. AD subjects on the ADNI dataset on the VQVAE generated data strongly preserve the known temporal lobe and hippocampal atrophy patterns associated with AD, producing a VBM t-map that strongly resembles the one from real data. Conversely, the VAE fails to show coherent structural differences in the GM.

Furthermore, we compared the volumes of key brain regions between populations of real and synthetic data. All images were segmented using GIF \cite{gif}, a robust multi-atlas based probabilistic segmentation model of the human brain which segments the brain into non-overlapping hierarchical 155 regions. Based on probabilistic segmentations, the total volume of each tissue was calculated, and then we ran a two-sided t-test to assess if there was a statistically significant difference between the tissue volumes of the real vs. synthetic populations. The Bonferroni-corrected target p-value was 2.083e-05.

\begin{table}[h]
\centering
\resizebox{1\columnwidth}{!}{%
\begin{tabular}{ccccccc}
Model         & Dataset & Population           & Gray Matter            & White Matter           & CSF                    & Deep Gray Matter     \\ \hline
Real          & UKB     & Young                & $595_{\pm32}$          & $460_{\pm29}$          & $280_{\pm21}$          & $40_{\pm3}$          \\
VQ-VAE (ours) & UKB     & Young                & $\mathbf{587_{\pm24}}$ & $\mathbf{472_{\pm20}}$ & $\mathbf{283_{\pm11}}$ & $\mathbf{40_{\pm2}}$ \\
VAE           & UKB     & Young                & $576_{\pm1}$           & $444_{\pm1}$           & $234_{\pm1}$           & $34_{\pm0}$          \\ \hline
Real          & UKB     & Old                  & $587_{\pm31}$          & $457_{\pm29}$          & $283_{\pm22}$          & $40_{\pm3}$          \\
VQ-VAE (ours) & UKB     & Old                  & $\mathbf{576_{\pm22}}$ & $\mathbf{465_{\pm20}}$ & $310_{\pm14}$          & $\mathbf{39_{\pm2}}$ \\
VAE           & UKB     & Old                  & $560_{\pm1}$           & $434_{\pm1}$           & $250_{\pm1}$           & $33_{\pm0}$          \\ \hline
Real          & UKB     & Small Ventricles     & $596_{\pm30}$          & $462_{\pm27}$          & $270_{\pm17}$          & $41_{\pm3}$          \\
VQ-VAE (ours) & UKB     & Small Ventricles     & $\mathbf{594_{\pm19}}$ & $477_{\pm18}$          & $280_{\pm12}$          & $\mathbf{41_{\pm2}}$ \\
VAE           & UKB     & Small Ventricles     & $572_{\pm1}$           & $444_{\pm1}$           & $235_{\pm1}$           & $35_{\pm0}$          \\ \hline
Real          & UKB     & Big Ventricles       & $589_{\pm34}$          & $459_{\pm30}$          & $283_{\pm19}$          & $41_{\pm3}$          \\
VQ-VAE (ours) & UKB     & Big Ventricles       & $574_{\pm20}$          & $\mathbf{467_{\pm17}}$ & $307_{\pm15}$          & $39_{\pm2}$          \\
VAE           & UKB     & Big Ventricles       & $570_{\pm1}$           & $442_{\pm1}$           & $246_{\pm1}$           & $34_{\pm0}$          \\ \hline
Real          & ADNI    & Cognitively Normal   & $530_{\pm51}$          & $430_{\pm40}$          & $309_{\pm32}$          & $40_{\pm5}$          \\
VQ-VAE (ours) & ADNI    & Cognitively Normal   & $\mathbf{554_{\pm19}}$ & $\mathbf{458_{\pm18}}$ & $\mathbf{299_{\pm12}}$ & $\mathbf{39_{\pm2}}$ \\
VAE           & ADNI    & Cognitively Normal   & $\mathbf{518_{\pm5}}$  & $\mathbf{440_{\pm4}}$  & $258_{\pm3}$           & $34_{\pm1}$          \\ \hline
Real          & ADNI    & Alzheimer's Diseased & $526_{\pm47}$          & $443_{\pm36}$          & $330_{\pm28}$          & $38_{\pm3}$          \\
VQ-VAE (ours) & ADNI    & Alzheimer's Diseased & $\mathbf{532_{\pm38}}$ & $\mathbf{446_{\pm20}}$ & $298_{\pm27}$          & $\mathbf{37_{\pm3}}$ \\
VAE           & ADNI    & Alzheimer's Diseased & $\mathbf{510_{\pm5}}$  & $\mathbf{443_{\pm4}}$  & $269_{\pm4}$           & $34_{\pm1}$          \\ \hline
\end{tabular}%
}
\caption{Tissue volumes based on GIF's probabilistic tissue segmentations. Mean and standard deviations were rounded to the nearest $10^3$. The bold values indicate the two-sided t-tests did not pass the statistical significance threshold compared to the real data.}
\label{tab:gif_table}
\end{table}
Overall, no significant volume differences were found between real and VQ-VAE samples for most subgroups and tissue types, while significant differences were found for most VAE statistics, demonstrating that the proposed method strongly preserves tissue volumes. The CSF volumes of the VQ-VAE UKB small and big ventricle populations were found to be statistically significantly different from their real counterparts as shown in Table \ref{tab:gif_table}, following the VBM results from Fig. \ref{fig:vbm}, and which could explain the increase in the t-statistic observed in the ventricular regions of the synthetic samples. On the other hand, the GM volumes were not statistically significantly different, corroborating the idea that the synthetic t-statistics are closer in magnitude to the real ones. Note that the VAE samples were also found not to be statistically significant in the ADNI AD/CT subset, but this is primarily due to the larger variance and the conservative Bonferroni correction. 


\section{Conclusion}
In this work, we propose a scalable and high-resolution volumetric generative model of the brain that preserves morphology. VBM \cite{vbm} and GIF \cite{gif} were used to assess the morphological preservation, while FID \cite{fid} and $bMMD^2$ \cite{mmd} to measure distribution alignment between synthetic and real samples. We have shown that the synthetic samples preserve healthy and pathological morphology and that they are realistic images that closely align with the distribution of the real samples. Future work should address the lack of conditioning and the top 1\% pruning to increase diversity and provide sampling control. Furthermore, the generative model could be extended for disease progression modelling, disentanglement of style and content, have its privacy preserving capabilities examined, and scaled to include multiple pathologies. To the best of our knowledge, this is the first morphologically preserving generative model of the brain, which paves the way for an unlimited amount of clinically viable data without jeopardizing patient privacy.

\subsubsection{Acknowledgements}

WHLP, MG, PB, MJC and PN are supported by Wellcome [WT213038/Z/18/Z].PTD is supported by the EPSRC Research Council, part of the EPSRC DTP [EP/R513064/1]. FV is supported by Wellcome/ EPSRC Centre for Medical Engineering [WT203148/Z/16/Z], Wellcome Flagship Programme [WT213038/Z/18/Z], The London AI Centre for Value-based Healthcare and GE Healthcare. PB is also supported by Wellcome Flagship Programme [WT213038/Z/18/Z] and Wellcome EPSRC CME [WT203148/Z/16/Z]. PN is also supported by the UCLH NIHR Biomedical Research Centre. The models in this work were trained on NVIDIA Cambridge-1, the UK’s largest supercomputer, aimed at accelerating digital biology.

\section{Appendix}

\subsection{VQ-VAEs} The VQ-VAE model has a similar architecture with \cite{walter_anomaly} but in 3D. The encoder uses strided convolutions with stride 2 and kernel size 4. There are four downsamplings in this VQ-VAE, giving the downsampling factor $f=2^4$. After the downsampling layers, there are three residuals blocks ($3\times3\times3$ Conv, ReLU, 1x1x1 Conv, ReLU). The decoder mirrors the encoder and uses transposed convolutions with stride 2 and kernel size 4. All convolution layers have 256 kernels. The $\beta$ in Eq. 1 is 0.25 and the $\gamma$ in Eq. 2 is 0.5. The codebook size was 2048 while each element's size was 32. \newline

\subsection{Transformers} Performer's\footnote{Implementation used: https://github.com/lucidrains/performer-pytorch} \cite{performer} has $L=24$ layers, $d=256$ embedding size, 16 multi-head attention modules (8 are local attention heads with window size of 420), and ReZero gating \cite{rezero}. Before the raster style ordering input was RAS+ canonical voxel representation oriented. \newline    

\subsection{Losses} VQ-VAE's pixel-space loss weight is 1.0, perceptual loss' weight is 0.001, frequency loss' weight is 1.0. The LPIPS uses AlexNet. Adam has been used as optimizer with an exponential decay of 0.99999. VQ-VAE's learning rate was 0.000165, discriminator's learning rate was 0.00005 and Performer's CrossEntropy learning rate was 0.001. \newline 

\subsection{Datasets} All datasets have been split into training and testing sub-sets. The VQ-VAE UKB sub-sets had 31740 and 3970 subjects respectively, while VQ-VAE ADNI had 648 and 82.  All datasets have been first processed with a rigid body registration such that they roughly fit the same field of view. Afterwards, all samples are passed through the following transformations before being fed into the VQ-VAE during training: first, they are being normalized to [0,1], then tightly spatially cropped resulting in an image of size (160,224,160), random affine (rotation range 0.04, translation range 2, scale range 0.05), random contrast adjustment (gamma [0.99, 1.01]), random intensity shift (offsets [0.0,0.05]), random Gaussian noise (mean 0.0, standard deviation 0.02), and finally, the images were thresholded to be in the range [0, 1.0]. For the Transformer, the UKB and ADNI datasets were split into sub-populations. UKB was split into small ventricles (6388 and 108), big ventricles (6321 and 156), young (6633 and 113), old (5137 and 106), while ADNI was split into cognitively normal (118 and 29) and Alzheimer's disease (151 and 36). For the Transformer training, each ADNI sample has been augmented 100 times and each augmentation's index-based representation was used for training it. \newline

\subsection{VBM analysis} For the Voxel-Based Morphometry (VBM), Statistical Parametric Mapping(SPM) \cite{spm12} package version 12.7486 was used with MATLAB R2019a. Before running the statistical tests, the images must first undergo unified segmentation where they were spatially normalized to a common template and simultaneously segmented into the  Gray Matter (GM), White Matter (WM), and Cerebrospinal fluid (CSF) tissue segments based on prior probability maps and voxel intensities. The unified segmentation was done with the default parameters: Bias Regularisation (light regularisation 0.001), Bias FWHM (60mm cutoff), MRF Parameter (1), Clean Up (Light Clean), Warping Regularisation ([0, 0.001, 0.5, 0.05, 0.2]), Affine Regularisation (ICBM space template - European brains), Smoothness (0), Sampling Distance (3). As per standard practice when using VBM, the group-aligned segmentations were modulated to preserve tissue volume, and a smoothing kernel was applied to the modulated tissue compartments to make the data conform to the Gaussian field model that underlines VBM and to increase the sensitivity to detect structural changes. The smoothing was also done with the default parameters with FWHM ([8, 8, 8]). For the VBM analysis, a Two-sample t-test Design was used, with the following parameters: Independence (Yes), Variance (Unequal), Grand mean scaling (No) and ANCOVA (No). No covariates, masking or global normalisation have been used. \newline

\noindent \textit{Appendix F - Additional Samples}
\begin{figure}[]
\centering
\includegraphics[width=1\columnwidth]{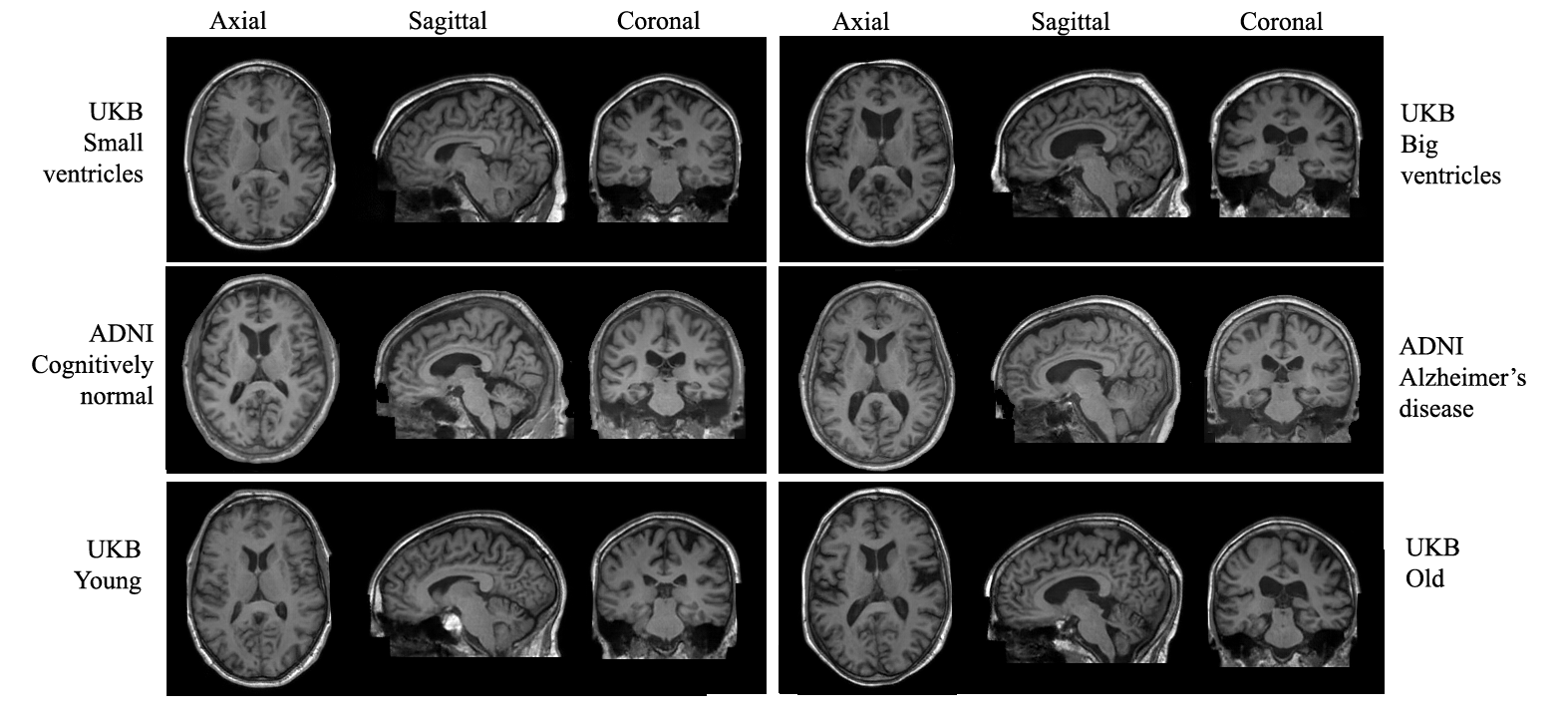}
\caption{Synthetic samples} 
\label{fig:syn}
\end{figure}

\printbibliography

\end{document}